\documentclass[aps,prl,twocolumn,showpacs,amsmath,amssymb,nofootinbib, preprintnumbers]{revtex4-1} 

\usepackage{color}

\newcommand{\be}{\begin{equation}}
\newcommand{\ee}{\end{equation}}
\newcommand{\ba}{\begin{eqnarray}}
\newcommand{\ea}{\end{eqnarray}}

\usepackage[pdftex]{graphicx} 

\def\ben{\begin{equation}}
\def\een{\end{equation}}
\def\half{\frac{1}{2}}
\def\bea{\begin{eqnarray}}
\def\eea{\end{eqnarray}}

\begin{document}
\newcount\hour \newcount\minute
\hour=\time  \divide \hour by 60
\minute=\time
\loop \ifnum \minute > 59 \advance \minute by -60 \repeat
\def\nowtwelve{\ifnum \hour<13 \number\hour:
                      \ifnum \minute<10 0\fi
                      \number\minute
                      \ifnum \hour<12 \ A.M.\else \ P.M.\fi
	 \else \advance \hour by -12 \number\hour:
                      \ifnum \minute<10 0\fi
                      \number\minute \ P.M.\fi}
\def\nowtwentyfour{\ifnum \hour<10 0\fi
		\number\hour:
         	\ifnum \minute<10 0\fi
         	\number\minute}
\def \now {\nowtwelve}

\begin{flushright}
\hfill UPR-1272-T
\end{flushright}

\title{An analytical formula for the vacuum polarization of rotating black holes}

\author{Mirjam Cveti\v c}
 \affiliation{Langberg Professor of Physics
University of Pennsylvania
Philadelphia, PA 19104-6396}
 \author{Zain H. Saleem}
\affiliation{Department of Physics and Astronomy,
 University of Pennsylvania, Philadelphia, PA 19104, USA\\
 \& National Center for Physics, Quaid-i-Azam university, Shahdara Valley Road, Islamabad, Pakistan}
\author{Alejandro Satz}
\affiliation{ Department of Physics and Astronomye
 University of Pennsylvania, Philadelphia, PA 19104, USA}
\date{\today}

\begin{abstract}
We give an analytical formula for the vacuum polarization of a massless minimally coupled scalar field at the horizon of a rotating black hole with subtracted geometry. This is the first example of an exact, analytical result for a four-dimensional rotating black hole.
\end{abstract}

\preprint{}

\maketitle

Quantum field theory in curved spacetime can be used to understand a lot of interesting features of black holes in a semiclassical approximation, most notably particle production near the black hole horizon  \cite{Hawking:1974sw}. The calculation of vacuum polarization or $\langle\phi^2\rangle$ (for a scalar field) is the simplest standard probe of quantum fluctuations in a black hole background, and can also be used to understand the symmetry breaking and Casimir effects near a black hole. Computation of $\langle\phi^2\rangle$ is also a preliminary step in evaluating the stress energy tensor $\langle T_{\mu\nu}\rangle$, which contributes to the backreaction through the semiclassical Einstein equation.

Candelas studied the vacuum polarization of a scalar field in the Schwarzschild black hole \cite{Candelas:1980zt} and was able to obtain an analytical expression for $\langle\phi^2\rangle $ at the horizon. Candelas' methods extend easily to charged static black holes; there have also been numerical  studies of vacuum polarization of scalar fields on general static black hole backgrounds beyond the event horizon (e.g. \cite{Anderson:1990jh} for asymptotically flat solutions and \cite{Flachi:2008sr} for the asymptotically anti-de Sitter case), and analytical computations at the horizon of a black hole threaded with a cosmic string \cite{Ottewill:2010hr}. The case of rotating black holes is much more challenging. Frolov \cite{Frolov:1982pi} was able to calculate the analytical expression for $\langle\phi^2\rangle $ only at the pole ($\theta=0$) of the event horizon, and Ottewill and Duffy \cite{Duffy:2005mz} have provided a numerical evaluation throughout the black hole horizon. However so far no one has been able to give an analytical formula for $\langle\phi^2\rangle$ throughout the horizon of a four-dimensional rotating black hole. (An analytic approximation good for fields with large mass is available, however \cite{Belokogne:2014ysa}, and exact results are obtainable in $d=3$ with AdS asymptotics \cite{Krishnan,Louko}.)

In this case we will be studying a particular example of rotating black holes that exist in "subtracted geometry" \cite{Cvetic:2011hp,Cvetic:2011dn,Cvetic:2012tr,Cvetic:2014sxa}. Subtracted geometry black holes are non extremal solutions of the bosonic sector of N=2 STU supergravity coupled to three vector multiplets. These black holes are obtained by subtracting some terms in the ``warp factor'' of the original black hole metric in such a way that the wave equation for a massless minimally coupled scalar field becomes separable and analytical solutions are obtainable. This subtracted black hole metric effectively places the black hole in an asymptotically conical box and mimics the ``hidden conformal symmetry'' \cite{Castro:2010fd} of the wave equation on rotating black holes in the near-horizon, near-extremal, and/or low energy regimes, which is a key motivator for the Kerr/CFT conjecture (see e.g. \cite{Compere:2012jk}). The energy density of the matter fields in this new geometry falls off as second power of radial distance, thus confining thermal radiation. The classical near horizon properties of the subtracted black hole are the same as the original black hole ones; in particular, the classical thermodynamics of the subtracted black hole is analogous to the standard one \cite{Cvetic:2014nta}, 
although loop corrections to the horizon entropy differ \cite{Cvetic:2014tka}). 

The horizon vacuum polarization in the static subtracted metric was studied in \cite{Cvetic:2014eka}. In this letter we shalll consider the subtracted geometry of the uncharged rotating Kerr black hole.  We shall see that the special features of the subtracted rotating metric, in particular the well-defined nature of the thermal vacuum and the solvability of the wave equation, allow us to obtain analytical results that are unavailable for the 
standard Kerr black hole.

The subtracted Kerr metric is given by:

\begin{align}
d  s^2  &= -  \Delta^{-1/2}  G \, 
( d{ t}+{ {\cal  A\, \mathrm{d}\tilde{\varphi}}})^2 \nonumber\\
&+ { \Delta}^{1/2}
\left(\frac{d r^2} { X} + 
d\theta^2 + \frac{ X}{  G} \sin^2\theta\, d\tilde{\varphi}^2 \right)\,.\label{metric4d}
\end{align}
with
\begin{eqnarray}
{ X} & =& { r}^2 - 2{ M }{ r} + { a}^2~,\;\;\; { G}  = { r}^2 - 2{ M}{ r} + { a}^2 \cos^2\theta\, \cr
{ {\cal A}}  &=&{2{ M} { a}r \sin^2\theta \over { G}}, \;\;\;\;  \Delta = 8 M^3 r  - 4M^2  a^2 \cos^2 \theta\,.
\end{eqnarray}
(The only difference between this metric and the standard Kerr metric is the form of the ``warp factor'' $\Delta$. For the  explicit form of gauge potentials and axio-dilatons of the STU model, supporting this geometry, see \cite{Cvetic:2012tr}.) The horizons and their surface gravities and angular velocities are given by:
\begin{eqnarray}\label{param}
r_\pm &=& M\pm \sqrt{M^2-a^2}\,, \cr
\kappa_\pm &=& \frac{1}{2M}\left[\frac{M}{\sqrt{M^2-a^2}}\pm1\right]^{-1}\,,\cr
\Omega_\pm &=&\kappa_\pm\frac{a} {\sqrt{M^2-a^2}} \,.
\end{eqnarray}

We switch to co-rotating coordinates $(t,r,\theta,\varphi)$, with the new angular variable being defined by:
\begin{equation}
\varphi=\tilde{\varphi} - \Omega_+ t\,.
\end{equation}
These are adapted to observers co-rotating with the black hole at the horizon. 
A noteworthy feature of subtracted geometry is that outside the horizon there is a globally defined timelike Killing vector, written as $\partial_t$ in the co-rotating coordinates \cite{Cvetic:2013lfa,Cvetic:2014ina}. This guarantees that there are no superradiant modes and ensures the existence of a Hartle-Hawking-like vacuum state adapted to the co-rotating observers. This is different from the case of ordinary Kerr black hole, where there is no such Killing vector \cite{kaywald,Ottewill:2000qh} and a physical co-rotating vacuum requires enclosing the black hole in a reflective box \cite{Frolov:1989jh,Duffy:2005mz}. The subtracted Kerr resembles more in this respect the Kerr/AdS black hole \cite{Krishnan}.

The general algorithm we follow for computing the horizon vacuum polarization in the Hartle-Hawking state starts by defining the Euclidean Green's function $G_H(x,x')$ (in a state regular at the horizon and infinity, and where the modes are adapted to co-rotating coordinates). Then we will evaluate $-iG_H$ with radial point splitting, perform the mode sum, and subtract the covariant divergent counterterms.

After writing the metric in coordinates $(t,r,\theta,\varphi)$ we perform the Wick rotation setting $t=-i \tau$. The metric becomes:
\begin{align}
d&s_E^2  = - \frac{G}{\Delta^{1/2} } \, 
[{ {\cal  A\, \mathrm{d}\varphi}}-i (1+{\cal A}\Omega_+)d\tau]^2  \nonumber\\
&+{ \Delta}^{1/2}
\left(\frac{d r^2} { X} + 
d\theta^2 + \frac{ X}{  G} \sin^2\theta\, (d\varphi-i\Omega_+ d\tau)^2 \right)\,.\label{metric4d}
\end{align}

 On writing the massless minimally coupled wave equation and proposing a solution of the form $\mathrm{e}^{i n\kappa_+ \tau}\mathrm{e}^{im\varphi}P_l^m(\cos\theta)\chi_{lmn}(r)$, we obtain straightforwardly a radial equation which, in the re-scaled variable $x =(r - \frac{1}{2}(r_+ + r_-))/( r_+ - r_-)$, reads:
\begin{align}
\,&\Big[ \frac{\partial}{\partial x} \left(x^2 - 
\frac{1}{4}\right)\frac{\partial}{\partial x}-    
\frac{n^2 }{4\left(x-\half \right)}\nonumber\\
&  + \frac{\beta_{mn}}{4\left( x+\half\right)}
 -  l(l+1) \Big]\, \chi_{lmn}(x)= 0\,, \label{wavehom}
\end{align}
where
\begin{equation}
\beta_{mn} = \frac{2 M n^2 r_-  - \,
 a^2 (4 m^2 + n^2)-4 i {a}  {m} {n} r_- }{r_+^2}\,.
\end{equation}

Two independent solutions of the equation, respectively regular at the horizon and at infinity, are:
\begin{equation}
\chi_{lmn}^{(1,2)} = \frac{\left( x-\half \right)^{\frac{n}{2}}}{\left(x+\half \right)^{\frac{n}{2 }+l+1}}
F \left(a_{lmn},b_{lmn}, c_{ln}^{(1,2)}
;z^{(1,2)}\right) \,,
\end{equation}
where 
\begin{align}
\,&c_{ln}^{(1)} = n+1\,,\,\, c_{ln}^{(2)} = 2l+2\,,\,\,z^{(1)}=\frac{x-\half}{x+\half}\,,\,\,z^{(2)}=\frac{1}{x+\half}\,,\nonumber\\
&(a_{lmn},b_{lmn})=l+1+\frac{|n|}{2}\pm\frac{\sqrt{\beta_{mn}}}{2}\,,
\end{align}
and the symmetry of the hypergeometric function makes irrelevant which branch of the square root is chosen. 

The full Green's function  is expanded as
\begin{align}\label{greenexpandrot}
\,&G_H( -i\tau , x , \theta , \varphi \; ; {-i \tau}' , {x}' , {\theta}' , {\varphi}')= \frac{ i \kappa}{ 2 \pi\,r_0} \sum_{n =- \infty}^ {\infty} e^{ i n \kappa( \tau -{\tau}')} \nonumber\\
&\sum_{l=0}^{\infty}
  \sum_{m =- l}^ {l} Y_l^m(\theta,\varphi)Y_l^{m*}(\theta',\varphi') G _{mln}(x,x')\,,
\end{align} 
where $r_0=r_+-r_-=2\sqrt{M^2-a^2}$, $\kappa\equiv \kappa_+$ as defined in (\ref{param}), and
\begin{equation}
G_{mln} ( x,x') = \frac{ \Gamma( a_{mln}) \Gamma(b_{mln})} {\Gamma(2l+2)\Gamma\left(1 +  \left| n  \right| \right) }\chi_{mln}^{(1)}( x_<) \chi_{mln}^{(2)}( x_>) \,.
\end{equation}
To evaluate the vacuum polarization at the horizon we set $x=1/2, x'=\epsilon+\frac{1}{2}$ (note that this is a dimensionless regulator $\epsilon=(r'-r)/r_0$) and join the points in the other directions, calling the resulting Green's function $G_H(\epsilon,\theta)$. All the terms in the sum vanish except $n=0$, so we are reduced to:
\begin{align}\label{greensum}
-&iG_H(\epsilon,\theta)= \frac{  \kappa}{ 8 \pi^2\,r_0}  \sum_{l=0}^{\infty}\sum_{m =- l}^ {l} \frac{(l-m)!}{(l+m)!}\left[P_l^m( \cos{\theta})\right]^2 \nonumber\\
&\times \frac{ \Gamma( l+1+i\alpha m) \Gamma( l+1-i\alpha m) } {\Gamma(2l+1)} \left(1+\epsilon\right)^{-(l+1)}\nonumber\\
&\times F\left(l+1+i\alpha m,l+1-i\alpha m,2l+2,\frac{1}{1+\epsilon}\right)\,,
\end{align} 
where the parameter  $\alpha\equiv a/r_+$ takes values between 0 and 1. 
We replace the hypergeometric by an integral expression using formula 9.111 of \cite{Grad}, leading to:
\begin{align}
-&iG_H(\epsilon,\theta)
= \frac{  \kappa}{ 8 \pi^2\,r_0}  \sum_{l=0}^{\infty}{(2l+1)}\sum_{m =- l}^ {l} \frac{(l-m)!}{(l+m)!}\left[P_l^m( \cos{\theta})\right]^2 \nonumber\\
&\times \int_0^1\mathrm{d}t\,\left(\frac{t(1-t)}{1+\epsilon -t}\right)^l \frac{1}{1+\epsilon-t} \cos\left(m \alpha \ln \lambda \right)\,,
\end{align} 
where $\lambda = \left(\frac{(1+\epsilon)(1-t)}{t(1+\epsilon-t)}\right)$.

The addition theorem for the associated Legendre polynomials is used to compute the sum over $m$, and formula III.4 from \cite{sansone} subsequently yields the sum over $l$. This leads, after a change of variables to $x=1-t$, to the integral expression
\begin{equation}
-iG_H(\epsilon,\theta)= \frac{  \kappa}{ 8 \pi^2\,r_0}   \int_0^1\mathrm{d}x\, f_\epsilon(x) \,;
\end{equation} 
\begin{equation}
  f_\epsilon(x)=\frac{\frac{\epsilon^2+2\epsilon x+(2-x)x^3}{(x^2+\epsilon)^3}}{\left[
1+\frac{4x(1-x)(x+\epsilon)}{(x^2+\epsilon)^2}\sin^2\theta \sin^2\left(\frac{\alpha}{2} \ln \lambda\right)
\right]^{3/2}}\,,
\end{equation} 
with $\lambda=\lambda(t(x))$.
It is easy to see from numerical evaluation that the leading divergences in the integral as $\epsilon\to 0$ match those provided by the standard counterterms \cite{Christensen:1976vb},
\begin{equation}\label{countersigma}
G_{div} = \frac{1+\frac{1}{12}R_{\mu\nu}\sigma^{,\mu}\sigma^{,\nu}}{8\pi^2\sigma} -\frac{1}{96\pi^2} R \ln (\mu^2 \sigma)\,,
\end{equation}
where $\sigma$ is the halved geodesic distance between the points and $\mu$ is an arbitrary mass scale. It is more difficult, however, to obtain an explicit expression for the finite result of the subtraction.
To make progress we perform the following sequence of changes of variables:
\begin{equation}
 u = \frac{1}{2}\ln\left(\frac{x(1+\epsilon)}{(1-x)(x+\epsilon)}\right)\,,\quad\quad w = \sinh u\,.
\end{equation}
This leads to the more tractable expression for the integral $I_\epsilon\equiv\int_0^1\mathrm{d}x\,f_\epsilon(x)$:
\begin{equation}
I_\epsilon =\int_0^\infty dw \frac{ \sqrt{1+\epsilon}}{\left[\epsilon + (1+\epsilon)w^2+v^2 \sin^2(\alpha \sinh^{-1} w)\right]^{3/2}}\,,
\end{equation}
where $v\equiv\sin\theta$. The intermediate $u$-integral expression is also obtainable directly from dimensional reduction from the Euclidean Green's function in AdS$^3\times$S$^2$, using the higher-dimensional embedding of subtracted geometry described in \cite{Cvetic:2011dn}\footnote{We thank Finn Larsen for bringing this point to our attention.}.

 To analyze the small $\epsilon$ limit and subtract explicitly the counterterms, we set aside momentarily the $\sqrt{1+\epsilon}$ prefactor and split the integral in two subintervals, $I_\epsilon^<$ over $(0, \epsilon^{1/6})$ and $I_\epsilon^>$ over $(\epsilon^{1/6},+\infty)$. In the second subinterval we can set $\epsilon$ to zero, at the expense of an error that vanishes as $\epsilon\to 0$. Then we can add and subtract terms compensating for the leading divergences at the lower limit, take $\epsilon\to 0$ safely in the subtraction, and integrate explicitly the added coutnerterms. This leads to: 
\begin{align}
I_\epsilon^>&\sim\int_{0}^\infty\mathrm{d}w\Bigg[\frac{1}{\left[w^2+v^2 \sin^2(\alpha \sinh^{-1} w)\right]^{3/2}} \nonumber\\
&-\left(\frac{1}{w^3(1+\alpha^2 v^2)^{3/2}}+\frac{v^2\alpha^2(1+\alpha^2)}{2w(1+w)(1+\alpha^2v^2)^{5/2}}\right)\Bigg]\nonumber\\
&+\frac{1}{2\epsilon^{1/3}(1+\alpha^2 v^2)^{3/2}}-\frac{v^2\alpha^2(1+\alpha^2)\ln \epsilon}{12(1+\alpha^2v^2)^{5/2}}\label{second}\,,
\end{align}
where $\sim$ stands for equivalence as $\epsilon\to 0$. The second subintegral is thus reduced to a finite integral involving no regulator, that can be evaluated numerically, plus two explicit divergent terms. 

In the first subinterval, we can show that:
\begin{align}
&I_\epsilon^<=\int_0^{\epsilon^{1/6}}\frac{\mathrm{d}w}{\left[\epsilon + (1+\epsilon)w^2+v^2 \sin^2(\alpha \sinh^{-1} w)\right]^{3/2}}\nonumber\\
&\sim\int_0^{\epsilon^{1/6}}\frac{\mathrm{d}w}{\left[\epsilon + (1+\epsilon)w^2+v^2\left(\alpha^2 w^2-\frac{\alpha^2(\alpha^2+1)w^4}{3}\right) \right]^{3/2}} \,,
\end{align}
which is expressible (formula 3.163.3 of \cite{Grad}) in terms of the incomplete elliptic integrals of first and second kind, $F(\gamma,k)$ and $E(\gamma,k)$. Here
\begin{equation}
\gamma = \arcsin\left(\frac{\epsilon^{1/6}}{\sqrt{c_+}}\sqrt{\frac{c_- + c_+}{c_- + \epsilon^{1/3}}}\right)\,,\quad\quad k = \sqrt{\frac{c_+}{c_-+c_+}}\,,
\end{equation}
and $c_{\pm}$ are the coefficients appearing in the denominator of the integrand when it is factored in a form proportional to $[(c_+^2-w^2)(c_-^2+w^2)]^{3/2}$.
We need the expansions of the elliptic functions near $(\gamma,k)= (\frac{\pi}{2},1)$, which have been derived in \cite{gustafson}. 
In order to obtain all the divergent and finite contributions to $I_\epsilon^<$, we need $F$ accurately to order $1$ and $E$ accurately to order $\epsilon$. This in turns require obtaining the argument $k$ accurately to order $\epsilon$ and $\gamma$ to order $\epsilon^{4/3}$. 
The result of this expansion is the following expression for the divergent and finite pieces of $I_\epsilon^<$:
\begin{align}\label{first}
I&_\epsilon^<\sim-\frac{1}{2\epsilon^{1/3}(1+\alpha^2v^2)^{3/2}}+\frac{1}{\epsilon\sqrt{1+\alpha^2v^2}}\nonumber\\
&+\frac{1}{6(1+\alpha^2v^2)^{5/2}}
\times\Big(-3-\alpha^2(7+4\alpha^2)v^2\nonumber\\
&+\alpha^2(1+\alpha^2)v^2(\ln(8(1+\alpha^2v^2)^{3/2})-\ln\epsilon)\Big) \,.
\end{align}
There is an additional finite contribution coming from the prefactor $\sqrt{1+\epsilon}$ to the integral, which yields when expanded a $1/2$ multiplied by the coefficient of the linear divergence of the integral. The complete result is thus expressed as:
\begin{equation}
I_\epsilon=I_\epsilon^<+I_\epsilon^>+ \frac{1}{2\sqrt{1+\alpha^2v^2}}\,,
\end{equation}
with the first to terms given by (\ref{first}) and (\ref{second}) respectively. We see that the $\epsilon^{-1/3}$ divergences cancel out, leaving only linear and logarithmic divergences that will match those of counterterms (\ref{countersigma}), leaving a finite renormalized result. 

This concludes the computation of the  explicit divergent and finite portions of the Green's function's coincidence limit. The counterterms (\ref{countersigma}) need now to be evaluated as a function of $\epsilon$ to the order $O(1)$. The form of $\sigma$ can be computed from the formulas expressing $\sigma$ in terms of coordinate separation:
\begin{align}
\sigma&=\frac{1}{2}g_{ab}\Delta x^a\Delta x^b + A_{abc}\Delta x^a\Delta x^b\Delta x^c\nonumber\\
&+ B_{abcd}\Delta x^a\Delta x^b\Delta x^c\Delta x^d+\cdots
\end{align}
where $A,B$ are obtained from symmetrized derivatives of the metric tensor, as described in \cite{Ottewill:2008uu}. 


These expressions are valid in a coordinate system in which the metric is regular. We use the Kruskal coordinates for the subtracted geometry that have been derived in \cite{Cvetic:2014ina}, which take the form $(U,V,\theta,\varphi)$ with $(-UV)\propto( r-r_+)$ near the horizon. Our radial coordinate separation is therefore written as $\Delta x^a = (-\delta,\delta,0,0)$ (with $\delta\propto\sqrt{\epsilon}$). After computing $\sigma$ by this procedure (leading to an expression of the form $\sigma=\beta_1\epsilon+\beta_2\epsilon^2+O(\epsilon^3)$) it is easy to obtain the Ricci counterterm in (\ref{countersigma}) because to the relevant order $O(\epsilon)$ we have $R_{\mu\nu}\sigma^{,\mu}\sigma^{,\nu} = R_{rr}\sigma^{,r}\sigma^{,r}$.



Once all the counterterms are computed by this procedure, when expressed in terms of the $\alpha$ parameter they take the relatively simple form:
\begin{align}
 &G_{div}=\frac{1+\alpha^2}{64\pi^2\,M^2}\Bigg[\frac{1}{\epsilon\sqrt{1+\alpha^2v^2}}-\frac{\alpha^2 v^2(1+\alpha^2)\ln \epsilon}{4(1+\alpha^2v^2)^{5/2}}\nonumber\\ &
+\frac{(-1+\alpha^2(-4+\alpha^2+(7+\alpha^2+\alpha^4)v^2+3\alpha^2v^4))}{12(1+\alpha^2v^2)^{5/2}} \Bigg]\,,
\end{align}
(plus a term of the form $R(r_+,\theta)\ln\mu^2$). Then, absorbing some $R$-proportional terms into the arbitrary constant $\mu$, the final result for the vacuum polarization is:
\begin{align}
\langle \phi^2\rangle&_{r_+}=R(r_+,\theta)\ln\mu^2+\frac{1+\alpha^2}{64\pi^2\,M^2}\Bigg\{\frac{1}{12(1+\alpha^2v^2)^{5/2}} \nonumber\\
\times&\Big[(1-\alpha^2(-4+\alpha^2(9+9\alpha^2+\alpha^4)v^2-3\alpha^2v^4))\nonumber\\
&-3\alpha^2(1+\alpha^2)^2\ln(1+\alpha^2v^2)\Big]\nonumber\\
&+\int_{0}^\infty\mathrm{d}w\Bigg[\frac{1}{\left[w^2+v^2 \sin^2(\alpha \sinh^{-1} w)\right]^{3/2}} \nonumber\\
&-\left(\frac{1}{w^3(1+\alpha^2 v^2)^{3/2}}+\frac{v^2\alpha^2(1+\alpha^2)}{2w(1+w)(1+\alpha^2v^2)^{5/2}}\right)\Bigg]\Bigg\}\,,
\end{align}
where
\begin{equation}
R(r_+,\theta)=\frac{3\alpha^2(1+\alpha^2)^2 v^2}{8\,M^2(1+\alpha^2v^2)^{5/2}}\,.
\end{equation}

\begin{figure}[h,t]
\begin{center}
\includegraphics[width=0.45\textwidth]{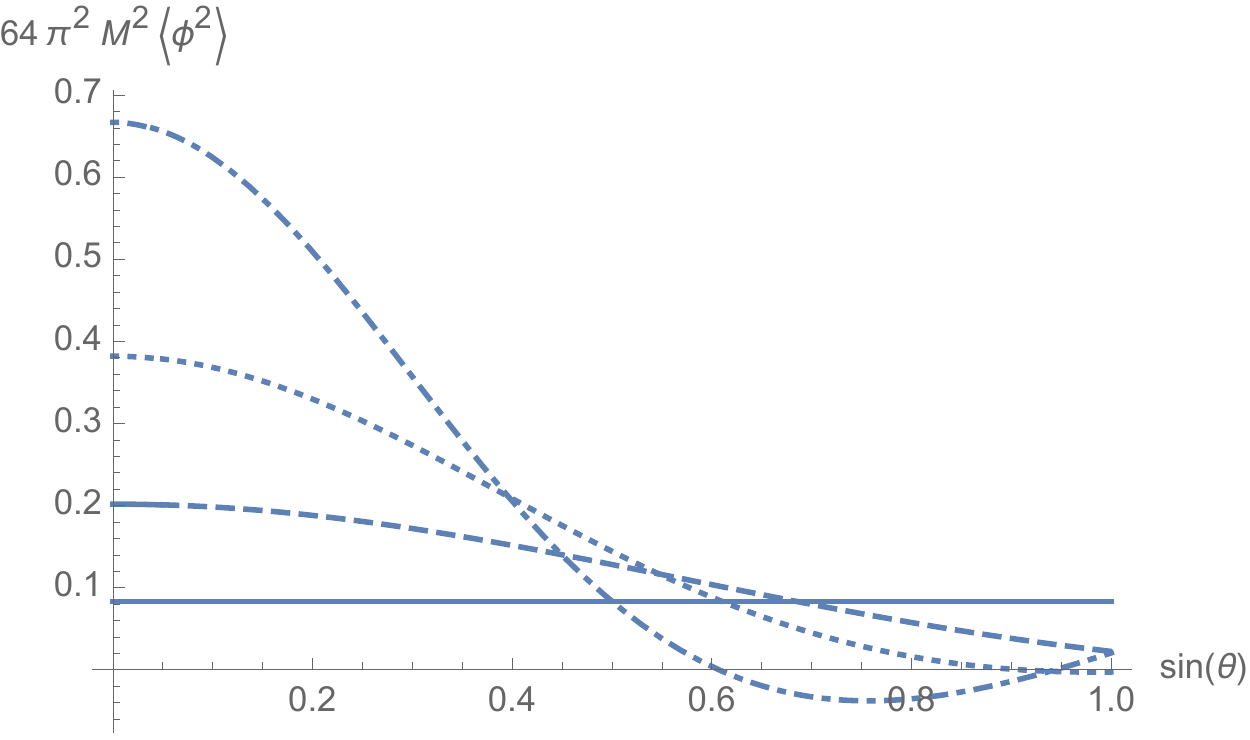}
\end{center}
\caption{Vacuum polarization (without $R$ term) at the horizon as a function of $v=\sin \theta$, for $\alpha\equiv{{a}/r_+} = 0$ (full line), $\alpha = 0.5$ (dashed line),
 $\alpha = 0.75$ (dotted line), and $\alpha = 1$ (dot-dashed line).}
\label{kerrpolots1}
\end{figure}

The angular profile for the vacuum polarization, neglecting the arbitrary term proportional to $R$,  is depicted in Figure 1. Notice that in the absence of rotation spherical symmetry is recovered, with its value  $\langle\phi^2\rangle_{r_+}^{Sch_{sub}}=(768\pi^2 M^2)^{-1}$ matching the result obtained in \cite{Cvetic:2014eka} for the subtracted Schwarzschild black hole. In addition, the result at the pole takes the form $\langle\phi^2\rangle_{r_+,\theta=0}=(768\pi^2 M^2)^{-1}(1+\alpha^2)(1+4\alpha^2-\alpha^4)$, agreeing with result found in  \cite{Cvetic:2014eka} using a non-corotating vacuum state (at the pole, the distinction is irrelevant). 
The dot-dashed plot corresponds to the extremal case $a = M$. 

It would be interesting to compare our results with numerical computations of the vacuum polarization in the standard Kerr metric (with a mirror in place to define the vacuum). Our calculation holds for the minimally coupled field, and the numerical results in \cite{Duffy:2005mz} are for the conformal case, so a direct comparison is not yet available. We expect our calculations to be easily generalized to the case of fields with higher spins as well as to rotating charged black holes, including multi-charged solutions \cite{Cvetic:1996kv,Cvetic:1996xz,Chow:2013tia}. We also expect our methods to be applicable to the computation 
of the stress-energy tensor, which would open the possibility of using the subtracted approximation to study analytically the backreaction for rotating four-dimensional black holes.

\vskip 0.1 in
\noindent{\bf Acknowledgements} 

We thank Finn Larsen and Gary Gibbons for valuable discussions and collaborations on related topics. MC would like to thank the  organizers of the  2015 Mitchell Institute Workshop at the Great Brampton House  for hospitality during the course of the work. The work is supported in part by the DOE (HEP) Award DE-SC0013528, the Fay R. and Eugene L. Langberg Endowed Chair (MC), and the Slovenian Research Agency (ARRS) (MC).

\providecommand{\href}[2]{#2}

\begingroup\raggedright

\end{document}